\RequirePackage{fix-cm}
\documentclass[twocolumn]{svjour3}          
\smartqed  
\usepackage{graphicx}
\usepackage{amsmath}
\usepackage{xcolor}
 \usepackage{mathptmx}      
\begin{document}

\title{Investigation of effects of pairing correlations on calculated $\beta$-decay half-lives of $fp$-shell nuclei}


\author{Asim Ullah \and Jameel-Un Nabi \and Muhammad Tahir}


\institute{Asim Ullah \at
              Faculty of Engineering Sciences, Ghulam Ishaq Khan Institute of Engineering Sciences and Technology,\\
              Topi, 23640, KP,
              Pakistan.
              \\
              \\
              \email{asimullah844@gmail.com}           
}

\date{Received: date / Accepted: date}

\maketitle
\begin{abstract}
Pairing of nucleons play a key role in solution to various nuclear physics problems. We investigate the probable effects of pairing correlations on the calculated Gamow-Teller (GT) strength distributions and the associated $\beta$-decay half-lives. Computations are performed for a total of 35 $fp$-shell nuclei using the proton-neutron quasiparticle random phase approximation (pn-QRPA) model. The nuclei were selected because of their importance in various astrophysical environments.  Pairing gaps are one of the key parameters in the pn-QRPA model to compute GT transitions. We employed three different values of the pairing gaps obtained from three different empirical formulae in our calculation. The GT strength distributions changed significantly as the pairing gap values changed. This in turn resulted in contrasting centroid and total strength values of the calculated GT distributions and led to differences in calculated half-lives using the three schemes. The half-life values computed via the three-term pairing formula, based on separation energies of nucleons, were in best agreement with the measured data. We conclude that the traditional choice of pairing gap values, $\Delta_p=\Delta_n={12/\sqrt A}$, may not lead to half-live values in good agreement with measured data. The findings of this study are interesting but warrants further investigation.

\end{abstract}
\section{Introduction}
\label{intro}
For open-shell nuclei, particle-hole methods (e.g. TDA and RPA) cannot be applied and the residual interaction becomes more important than in the pure particle–hole picture. The short-range part of the residual interaction manifests itself as nucleon pairing. The pairing interaction lowers the total energy by an amount $2\Delta$, where $\Delta$ is normally termed as the pairing gap. Yet another aspect of nucleon pairing is witnessed in the odd-even effect. Experiment has established that the total binding energy of an odd-A nucleus is less than the average of the total
binding energies of the two neighbouring even–even nuclei. In case of deformed nuclei, moments of inertia calculated without pairing are 2--3 times larger than the measured ones. 

Of particular importance is the role of nucleon pairing in nuclear astrophysics. The structure of a nucleus within the core of a massive star plays a crucial role in the process of a supernova explosion and determines the road-map for evolution of stars \cite{Bet79}. According to numerical simulations, the stellar evolution is primarily determined by the temporal variation of the lepton fraction (Y$_e$), which is governed by the weak interaction (especially electron capture and $\beta$-decay). Weak interactions are the key components in several stellar processes, including presupernova evolution, and nucleosynthesis. $\beta$-decay and electron capture play a significant
role during the core collapse of a massive star. To explore the mechanism of supernova (both Type-Ia and Type-II) explosions, the magnitude of the electron capture rate is the most critical factor \cite{Ful80,Ful82a,Lan03}. In a Type-II supernovae, once the mass of the Fe core exceeds the Chandrasekhar mass limit, the degenerate electron gas pressure can no longer withstand gravitational force, and the core begins to collapse. Electron capture (\textit{ec}), on the one hand, lowers the lepton fraction, which in turn reduces the electron degenerate pressure. On the other hand, the \textit{ec} (and also the $\beta$-decay) results in the production of (anti)neutrinos, which channel the energy away from the core and may accelerate the collapse. A Type-Ia supernova is considered to be the result of a thermonuclear explosion on an accreting white dwarf, and its collapse is thought to be the result of the general relativistic effect. However, the \textit{ec} process is believed to be responsible for the abundance of certain iron isotopes in Type-Ia supernovae \cite{Lan03}.

There are around $6000$ nuclei between the $\beta$ stability and the neutron drip line. Majority of these nuclei cannot be generated in terrestrial laboratories and hence their $\beta$-decay characteristics must be calculated theoretically.  The calculations of weak interaction rates under stellar conditions rely heavily on reliable computation of the ground and excited states Gamow-Teller (GT) response \cite{Bet79}. In atomic nuclei, GT transitions are the most common spin–isospin ($\sigma$$\tau$) type nuclear weak processes \cite{Ost92}. The isospin has three components in a spherical coordinate system: $\tau_+$, $\tau_-$ and $\tau_0$. The $\tau_{+}$ stands for GT transitions $({GT}_{+})$ in beta positive direction, whereas $\tau_{-}$ indicates GT transitions $({GT}_{-})$ in beta negative direction. The third component $\tau_{0}$ is important for inelastic neutrino-nucleus scattering at low neutrino energies. At the initial stage of core-collapse, when the temperatures and densities are low (300 - 800 KeV and $\sim$ 10$^{10}$ g/cm$^3$, respectively), the nuclear Q-value and electron chemical potential have nearly equivalent magnitudes. In such a scenario, the \textit{ec} rates are highly dependent on the details of the distribution of GT strength. The centroid and total GT strength values control the \textit{ec} rates when the chemical potential surpasses the Q-value at relatively high core-densities. That is why, a thorough understanding of the GT distributions is required for a reliable calculation of stellar rates and $\beta$-decay half-lives.

Numerous initiatives to examine the $\beta$-decay characteristics have been made in the past. Few important mentions are; the gross theory \cite{Tak73}, the proton-neutron quasiparticle random phase approximation (pn-QRPA) approach \cite{Nab99,Hir93} and the shell model techniques  \cite{Mar99}. To evaluate the $\beta$-decay characteristics, the gross theory follows a statistical recipe. The shell model and the pn-QRPA approaches, on the other hand, are microscopic in nature. Shell model (SM) results may be accurate only for light nuclei \cite{Tan17}. Moreover, the SM incorporates the contribution of excited state GT strength distributions at high temperatures using the Brink-Axel hypothesis \cite{Bri58}. The pn-QRPA can be used for both light and heavy nuclei to calculate their $\beta$-decay properties \cite{Wan16}. Furthermore, the exited state GT transitions can be computed using the pn-QRPA model without using the Brink's hypothesis.

The pairing gaps is one of the most important parameters employed in the pn-QRPA model for calculation of GT strength distributions and $\beta$-decay half-lives~\cite{Sta90,Hir93}. To cope with nucleon pairing effect, the BCS approach is applied.
In this study, we explore how pairing gaps affect the calculated GT strength distributions and associated decay half-lives. We calculated the $\beta$-decay half-lives and GT strength distributions of a total of 35 $fp$-nuclei using the pn-QRPA model. These nuclei were selected from a list of important weak interaction nuclei compiled recently by Nabi and collaborators \cite{Nab21}. The pairing gaps were computed via three different empirical formulae and would  be discussed in the next section. The calculated half-lives were later compared with recent experimental data \cite{Kon21}.  \\ 
The following is the outline of the paper. The formalism employed in our calculation is briefly described in Section 2. The third section discusses our findings. The last section includes a summary and conclusion.

\section{Formalism}
\label{sec:1}
The Hamiltonian of the pn-QRPA model was taken to be composed of four components:
\begin{equation} \label{H}
	H^{QRPA} = H^{sp} + V^{pair} + V^{pp}_{GT} + V^{ph}_{GT},
\end{equation}
where $H^{sp}$ stands for the single-particle Hamiltonian, $V_{GT}^{ph}$
and $V_{GT}^{pp}$ represent the particle-hole (\textit{ph}) and particle-particle (\textit{pp}) GT
forces, respectively. The last term $V^{pair}$ denotes the pairing
force which was computed under the BCS approximation. The last three terms in our Hamiltonian result from the residual interaction. The single-particle energies and wavefunction were computed using the Nilsson model \cite{Nil55}, which included nuclear deformation. The equation $\hbar\omega=41A^{1/3}$ was used to calculate the oscillator constant for nucleons. The Nilson-potential
parameter were adopted from Ref. \cite{Rag84}. $Q$-values were adopted from the recent compilation of Ref. \cite{Kon21}. The values of deformation parameter ($\beta_2$) were taken from Ref.~\cite{Mol16}. \\
The spherical nucleon basis ($c^{\dagger}_{jm}$, $c_{jm}$) was transformed to the deformed basis ($d^{\dagger}_{m\alpha}$, $d_{m\alpha}$) employing the following equation
\begin{equation}\label{df}
	d^{\dagger}_{m\alpha}=\Sigma_{j}D^{m\alpha}_{j}c^{\dagger}_{jm},
\end{equation}
where $c^{\dagger}$ ($d^{\dagger}$) is the particle creation operator in the spherical basis (deformed basis). The Nilsson Hamiltonian was diagonalized to get the transformation matrices ($D$) where  $\alpha$ represents additional quantum numbers. We performed separate BCS calculations for the neutron and proton systems. A pairing force of constant strength G ($G_n$ and $G_p$ for neutrons and protons, respectively) was adopted in the current calculation:
\begin{eqnarray}\label{pr}
	V_{pair}=-G\sum_{jmj^{'}m^{'}}(-1)^{l+j-m}c^{\dagger}_{jm}c^{\dagger}_{j-m}\\ \nonumber
	~~~~~~~~~~~~~~~~~(-1)^{j^{'}+l^{'}-m^{'}} c_{j^{'}-m^{'}}c_{j^{'}m^{'}}.
\end{eqnarray}
Later we introduced a quasiparticle basis $(a^{\dagger}_{m\alpha}, a_{m\alpha})$ employing the Bogoliubov transformation
\begin{equation}\label{qbas}
	a^{\dagger}_{m\alpha}=u_{m\alpha}d^{\dagger}_{m\alpha}-v_{m\alpha}d_{\bar{m}\alpha}
\end{equation}
\begin{equation}
	a^{\dagger}_{\bar{m}\alpha}=u_{m\alpha}d^{\dagger}_{\bar{m}\alpha}+v_{m\alpha}d_{m\alpha},
\end{equation}
where $\bar{m}$ is the time-reversed states of $m$ and $a$ ($a^{\dagger}$) stands for the quasiparticle (q.p.) annihilation (creation) operator, which appears in the RPA equation. The occupation amplitudes ($v_{m\alpha}$ and $u_{m\alpha}$) were calculated within the BCS approximation (subject to $v^{2}_{m\alpha}$ + $u^{2}_{m\alpha}$ = 1).\\
The GT transitions were expressed in terms of QRPA phonons:
\begin{equation}\label{co}
	A^{\dagger}_{\omega}(\mu)=\sum_{pn}[X^{pn}_{\omega}(\mu)a^{\dagger}_{p}a^{\dagger}_{\overline{n}}-Y^{pn}_{\omega}(\mu)a_{n}a_{\overline{p}}],
\end{equation}
where $p$ ($n$) stands for $m_{p}\alpha_{p}$ ($m_{n}\alpha_{n}$). The sum includes all proton-neutron pairs subject to the conditions $\mu=m_{p}-m_{n}$ and $\pi_{p}.\pi_{n}$=1, with $\pi$ denoting parity. The $X$ ($Y$) denotes the forward-going (backward-going) amplitude. The proton-neutron residual interactions in the pn-QRPA approach occur via \textit{pp} and \textit{ph} channels, defined by interaction constants $\kappa$ and $\chi$, respectively. The $ph$ GT force was determined using:
\begin{equation}\label{ph}
	V^{ph}= +2\chi\sum^{1}_{\mu= -1}(-1)^{\mu}Y_{\mu}Y^{\dagger}_{-\mu},\\
\end{equation}
with
\begin{equation}\label{y}
	Y_{\mu}= \sum_{j_{p}m_{p}j_{n}m_{n}}<j_{p}m_{p}\mid
	t_- ~\sigma_{\mu}\mid
	j_{n}m_{n}>c^{\dagger}_{j_{p}m_{p}}c_{j_{n}m_{n}},
\end{equation}
whereas the $pp$ GT force was computed using:
\begin{equation}\label{pp}
	V^{pp}= -2\kappa\sum^{1}_{\mu=
		-1}(-1)^{\mu}P^{\dagger}_{\mu}P_{-\mu},
\end{equation}
with
\begin{eqnarray}\label{p}
	P^{\dagger}_{\mu}= \sum_{j_{p}m_{p}j_{n}m_{n}}<j_{n}m_{n}\mid
	(t_- \sigma_{\mu})^{\dagger}\mid
	j_{p}m_{p}>\times \nonumber\\
	(-1)^{l_{n}+j_{n}-m_{n}}c^{\dagger}_{j_{p}m_{p}}c^{\dagger}_{j_{n}-m_{n}},
\end{eqnarray}
where the rest of the symbols have their traditional meanings. The \textit{ph} and \textit{pp} force have different signs revealing their opposite nature. The interaction strengths $\kappa$ and $\chi$ were chosen as $0.58/A^{0.7}$ and $5.2/A^{0.7}$, respectively, adopted from Ref.~\cite{Hom96}.  Our calculation satisfied the model independent Ikeda sum rule \cite{Ike63}. The reduced GT transition probabilities from the QRPA ground state to one-phonon states in the daughter nucleus were calculated as
\begin{equation}
	B_{GT} (\omega) = |\langle \omega, \mu ||\tau_{\pm} \sigma_{\mu}||QRPA \rangle|^2.
\end{equation}
We refer to~\cite{Hir93,Sta90} for details of full solution of Eq.~(\ref{H}). \\
The $\beta$-decay partial half-lives were computed via the following relation:
\begin{eqnarray}
	t_{p(1/2)} = \nonumber\\\frac{C}{(g_A/g_V)^2f_A(Z, A, E)B_{GT}(\omega)+f_V(Z, A, E)B_F(\omega)},
\end{eqnarray}
where $E$ = $Q$ - $\omega$, $\omega$ (energy of the final state), value of $g_A/g_V$ was taken as -1.254 \cite{War94} and C (= ${2\pi^3 \hbar^7 ln2}/{g^2_V m^5_ec^4}$)  was adopted as 6295 $s$. $f_A(E, Z, A)$ and $f_V(E, Z, A)$ are the phase space integrals for axial vector and vector transitions, respectively. $B_{F}$ ($B_{GT}$) stands for the reduced transition probability for the Fermi (GT) transitions.
Finally, the total $\beta$-decay half-lives were calculated using the equation
\begin{equation}
	T_{1/2} = \left(\sum_{0 \le E_j \le Q} \frac{1}{t_{p(1/2)}}\right)^{-1}.
\end{equation} 
The summation includes all the transition probabilities to the states in daughter within the $Q$ window.\\
As mentioned earlier, pairing gap values are key model parameters in the pn-QRPA approach. Three different values of pairing gaps were used in the current calculation in order to explore their impact on the computed $\beta$-decay half-lives and GT strength distributions. The first one was computed using the traditional and mass-dependant relation $\Delta_p=\Delta_n={12/\sqrt A}$ MeV (also supported by the liquid-drop model of the nucleus). The second recipe consists of three terms. It computes different pairing gaps for neutrons and protons. The relationship is expressed in terms of proton and neutron separation energies as:
\begin{eqnarray}
	\bigtriangleup_{pp} =\frac{1}{4}(-1)^{Z+1}[S_p(Z+1, A+1)\nonumber\\-2S_p(Z, A)+S_p(Z-1, A-1)]
\end{eqnarray}
\begin{eqnarray}
	\bigtriangleup_{nn} =\frac{1}{4}(-1)^{A-Z+1}[S_n(Z, A+1)\nonumber\\- 2S_n(Z, A) + S_n(Z, A-1)]
\end{eqnarray} 
The third formula consist of five binding energy terms and is given as:
\begin{eqnarray}
	\bigtriangleup_{nn} = \frac{1}{8}[B(N-2, Z) - 4B(N-1, Z) + 6B(N, Z) \nonumber\\- 4B(N+1, Z) + B(N+2, Z)]
\end{eqnarray}
\begin{eqnarray}
	\bigtriangleup_{pp} = \frac{1}{8}[B(N, Z-2) - 4B(N, Z-1) + 6B(N, Z)\nonumber\\ - 4B(N, Z+1) + B(N, Z+2)]
\end{eqnarray}
The values of binding energy were adopted from the recent atomic mass evaluation~\cite{Wan21}.
The first, second, and third schemes are referred to as TF, 3TF, and 5TF, respectively, for ease of reference.
\section{Results and Discussion}
A total of 35 $fp$-shell nuclei of astrophysical importance~\cite{Nab21} were short-listed for the current calculation. Out of the 35 selected nuclei, 17  decay via electron emission while 18 are unstable to $\beta^+$-decay. The $\beta$-decay half-lives and GT strength distributions of the selected nuclei were computed using the pn-QRPA approach. To investigate the impact of the pairing gaps, we employed three different values of the pairing gaps (TF, 3TF and 5TF) in our calculation. The computed $\beta$-decay half-lives were later compared with the measured data~\cite{Kon21}.\\
On the basis of pairing gap values, we may encounter four different cases. The 	$\bigtriangleup_{pp}$ values, calculated using the 3TF/5TF schemes, can be bigger (or smaller) as compared to the traditional choice of  $\bigtriangleup_{pp=nn}$ values of the TF scheme \textit{and} the 	$\bigtriangleup_{nn}$ values, calculated using the 3TF/5TF schemes, can be smaller (or bigger) as compared to the  $\bigtriangleup_{pp=nn}$ of the TF scheme. These two cases would be denoted by C1 and C2, respectively. On the other hand, it is also possible that \textit{both} $\bigtriangleup_{pp}$ and $\bigtriangleup_{nn}$ values of the 3TF/5TF schemes are bigger or smaller than the $\bigtriangleup_{pp=nn}$ values of the TF scheme. The later two cases would be referred to as C3 and C4, respectively, in this paper. It may be noted that we have fewer nuclei in category C2 and C3. As stated earlier, our criteria for choosing the 35 fp-shell nuclei were their astrophysical importance  (as per recent finding of Ref.~\cite{Nab21}) and an equal number of $\beta^+$ and $\beta^-$ cases. Nonetheless, all four categories are well represented in our chosen ensemble of nuclei. 

\begin{figure*}[]
	\centering
	\includegraphics[width=1\textwidth]{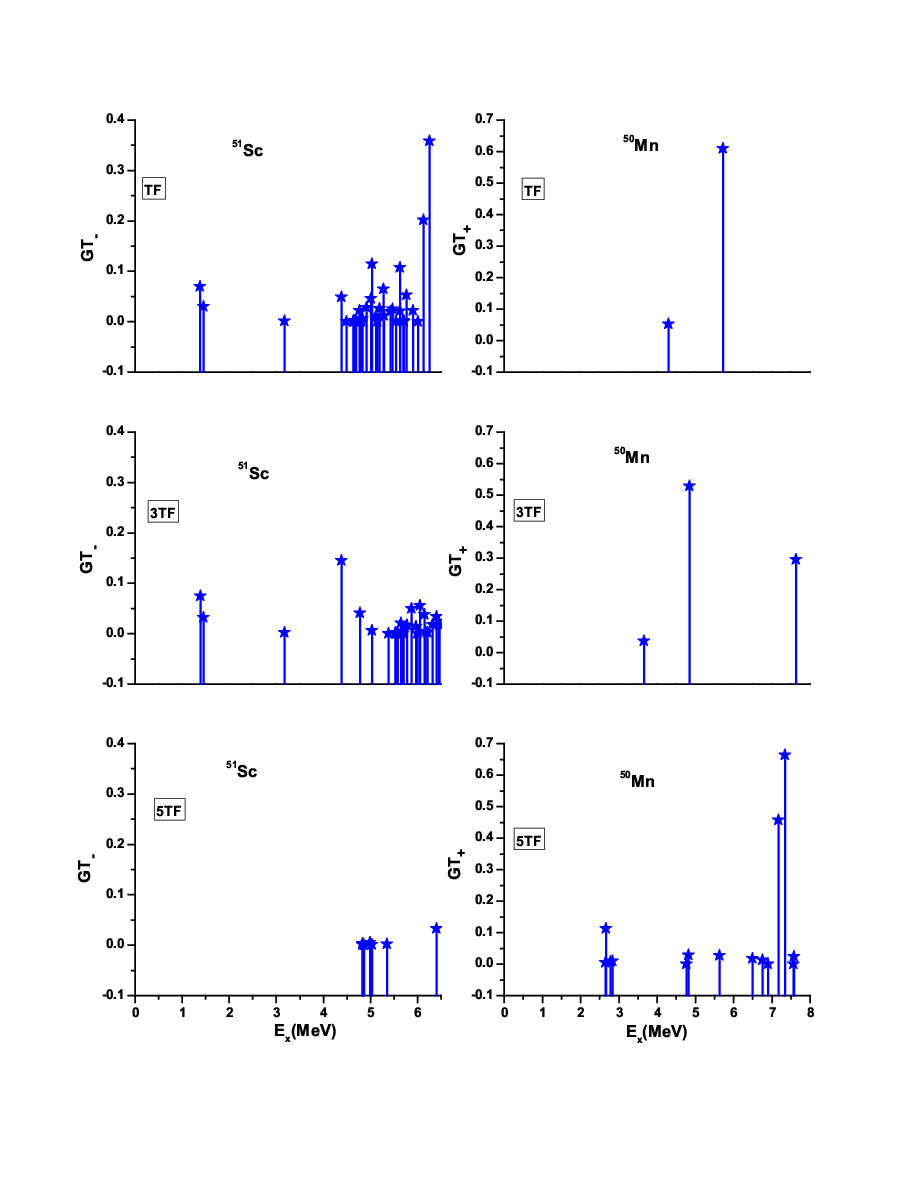}
	\vspace*{-2.5cm}
	\caption{Calculated GT strength distributions, in $\beta^-$ direction, for $^{51}$Sc and $^{50}$Mn as a function of daughter excitation energies using the three pairing gap values.\label{F1} }
\end{figure*}

\begin{figure*}[]
	\centering
	\includegraphics[width=1\textwidth]{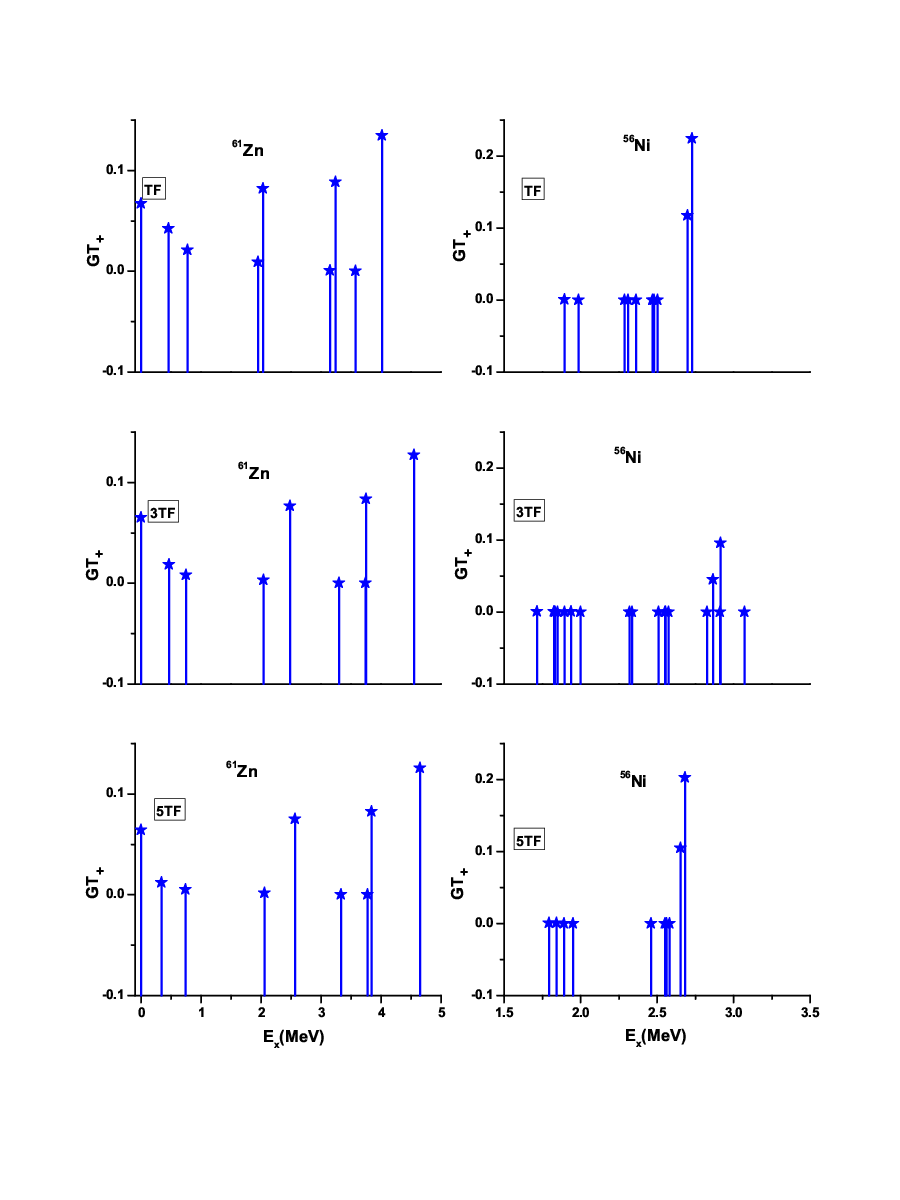}
	\vspace*{-2.5cm}
	\caption{Calculated GT strength distributions, in $\beta^+$ direction, for $^{61}$Zn and $^{56}$Ni as a function of daughter excitation energies using the three pairing gap values.\label{F2} }
\end{figure*}

The sample GT strength distributions for the case of $^{51}$Sc (C1), $^{61}$Zn (C2), $^{56}$Ni (C3) and $^{50}$Mn (C4), using the TF, 3TF and 5TF computed pairing gaps, are shown in Figs.~(\ref{F1}-\ref{F2}). All the three formulae led to different strength distributions (albeit less for the C2 case of $^{61}$Zn). In general, it is noted that the 3TF/5TF schemes result in more fragmentation of the GT strength (at times outside the Q-value window). The changes in the strength distributions altered the calculated total GT strength and centroid values. The calculated $\beta$-decay half-lives also changed which we discuss below. We further notice that a bigger pool of data would have been better for performing a statistical analysis.

The cumulative GT strength (in arbitrary units) and centroids (in MeV units) of the calculated GT strength distributions for the C1 \& C2 cases are presented in Table 1. It is noted that, in general, the TF scheme resulted in bigger total GT strength and smaller centroid values when compared with the corresponding values of the 3TF/5TF schemes.  
Table~2 shows the total GT strength and centroid values for the C3 \& C4 cases. We were unable to notice any systematic trend in the computed centroid values of the resulting GT strength distributions. However, in general, we did notice that the TF scheme computed smaller total GT strength values in C3 and C4 cases. It is further noted that the pn-QRPA model calculated GT transitions above the Q-value window for three cases in 5TF scheme and once instance in 3TF scheme (represented by dashes in the tables).

Table~3 shows the  pn-QRPA calculated half-life values employing the three different schemes (TF, 3TF and 5TF), for the C1 \& C2 cases, whereas Table~4 depicts calculated half-life values for C3 \& C4 cases. The calculated half-life values are compared with the experimental data adopted from Ref.~\cite{Kon21} and shown in the last column. The trends in the computed half-life values for the three schemes may be explained from the data of Tables~(1-2). Bigger values of total GT strength and lower values of GT centroid translated into smaller values of $\beta$-decay half-lives. We computed the standard deviation of calculated half-lives from the measured data for the three schemes. 
The lowest standard deviation of 102 $s$ was noted for the 3TF scheme, which was followed by a standard deviation of 762 $s$ for the 5TF scheme. The TF scheme resulted in biggest standard deviation of 6879 $s$. It is to be noted that we excluded the case of $\beta^+$-decay of $^{45}$Ti in our calculation of standard deviation for all the three schemes because of the missing entry in case of 5TF scheme.
It was further noted that the 3TF scheme reproduced 15 $\beta$-decay half-lives within a factor of 2 (the number of corresponding cases for TF and 5TF schemes were 11 and 10, respectively). We therefore conclude that pairing gaps computed as a function of separation energies of nucleons resulted in $\beta$-decay half-lives in better agreement with the measured data. However, we again remark that our investigation is in preliminary stages and might require some modifications as we increase the pool size of our data.

The branching ratios were calculated employing the following relation:
\begin{eqnarray}
	I = \frac{T_{1/2}}{t^{par}_{1/2}} \times 100 (\%)
\end{eqnarray}
where $T_{1/2}$ stands for the total half-life.
Tables (5-8) show the state-by-state GT strength, branching ratios (I) and partial half-lives ($t^{par}_{1/2}$) for the decay of $^{62}$Fe (C1 case), $^{62}$Zn (C2 case), $^{48}$Cr (C3 case) and $^{58}$Cu (C4 case), respectively. It is noted that the TF scheme resulted in lesser fragmentation of the GT strength when compared with the GT distributions of 3TF and 5TF schemes. The partial half-lives and state-by-state GT strength of all remaining  nuclei may be requested as ASCII files from the corresponding author.\\

\section{Summary and Conclusion} 
We explore the impact of pairing correlations on the calculated $\beta$-decay characteristics of the important $fp$-shell nuclei. $\beta$-decay half-lives and GT strength distributions for 35 important $fp$-shell nuclei (adopted from the list compiled by Ref.~\cite{Nab21}) were calculated using the pn-QRPA model. We included equal number of $\beta^+$  and $\beta^-$ decay cases in our investigation. Pairing gap values between the paired nucleons is one of the key model parameters in the pn-QRPA theory. Three different values of pairing gaps, computed using three different recipes, were used in the current investigation in order to study their impact on the calculated $\beta$-decay properties of the astrophysically important unstable nuclei. As expected, the GT strength distributions,  centroid values and half-lives were significantly altered as the pairing gap values were changed. For C1 and C2 cases, we concluded that the 3TF and 5TF schemes led to smaller calculated total GT strength and higher centroid values when compared with the GT distribution of the TF scheme. It was further noted that, in general, the TF scheme computed smaller GT strength values for C3 and C4 cases. For our selected pool of nuclei, consisting both of small and large half-life values, the three-term formula (3TF), based on neutron and proton separation energies, was found to match best the measured data. It is also remarked that a bigger pool of nuclei is required to substantiate the findings of the current investigation.
\section*{Acknowledgement}
J.-U. Nabi would like to acknowledge the support of the Higher Education Commission Pakistan through project
20-15394/NRPU/R\&D/HEC/2021.

\section*{Declaration of Competing Interest}
The authors declare that they have no known competing financial
interests or personal relationships that could have appeared to influence the work reported in this paper.

\begin{table*}[]\label{T1}
	\centering
	\caption{Calculated centroid and total GT strength values of the GT strength distributions for C1 (upper panel) \& C2 (lower panel) cases using the three computed pairing gaps (TF, 3TF and 5TF). }
	\centering
	\begin{tabular}{c|c|cc|cc|ccc|ccc}
		\hline
		\hline
		&\multicolumn{5}{c}{Pairing Gaps (MeV)}	&\multicolumn{3}{c}{Total Strength (arb. units)}&\multicolumn{3}{c}{Centroids (MeV)} \\
		\hline
		Nuclei&$\bigtriangleup^{TF}_{nn=pp}$&$\bigtriangleup^{3TF}_{pp}$&$\bigtriangleup^{3TF}_{nn}$&$\bigtriangleup^{5TF}_{pp}$&$\bigtriangleup^{5TF}_{nn}$&$\sum$ GT$^{TF}$ &$\sum$ GT$^{3TF}$ &$\sum$ GT$^{5TF}$ &$\bar{E}$$^{TF}$&$\bar{E}$$^{3TF}$&$\bar{E}$$^{5TF}$\\
		\hline
		\hline
		$^{51}$Sc & 1.680 & 2.221 & 0.545 & 1.913 & 0.574 & 1.302 & 0.607 & 0.045 & 5.364 & 4.753 & 6.037 \\
		$^{49}$Sc & 1.714 & 2.177 & 1.494 & 1.917 & 1.350 & 0.550 & 0.528 & ---   & 1.327 & 1.327 & ---   \\
		$^{57}$Cu & 1.589 & 2.016 & 1.502 & 1.650 & 1.332 & 1.396 & 1.277 & 0.624 & 6.214 & 6.161 & 6.480 \\
		$^{49}$Ca & 1.714 & 1.969 & 1.505 & 1.952 & 1.097 & 0.025 & 0.003 & 0.003 & 0.805 & 1.415 & 1.420 \\
		$^{65}$Co & 1.488 & 1.618 & 0.906 & 1.824 & 0.936 & 0.573 & 0.571 & 0.637 & 2.996 & 2.985 & 3.878 \\
		$^{63}$Co & 1.512 & 1.609 & 1.098 & 1.724 & 1.042 & 0.705 & 0.134 & 0.129 & 1.222 & 0.713 & 0.405 \\
		$^{62}$Fe & 1.524 & 1.616 & 1.412 & 1.612 & 1.428 & 0.272 & 0.690 & 0.691 & 1.234 & 1.435 & 1.436 \\
		$^{58}$Cr & 1.576 & 1.638 & 1.392 & 1.618 & 1.448 & 1.043 & 1.022 & 1.028 & 2.163 & 2.145 & 2.151 \\
		$^{59}$Cu & 1.562 & 1.610 & 0.759 & 1.623 & 0.924 & 0.351 & 0.110 & 0.119 & 2.966 & 3.358 & 3.299 \\
		$^{59}$Mn & 1.562 & 1.593 & 0.903 & 1.642 & 0.900 & 0.284 & 0.288 & 0.231 & 2.015 & 2.038 & 2.746 \\
		$^{57}$Mn & 1.589 & 1.607 & 0.902 & 1.686 & 0.900 & 0.235 & 0.236 & 0.102 & 0.923 & 0.922 & 1.041 \\
		$^{50}$Sc & 1.697 & 1.934 & 1.206 & 1.648 & 0.875 & 0.066 & 0.007 & 0.009 & 3.825 & 4.942 & 4.465 \\
		\hline
		$^{45}$Ti & 1.789 & 1.229 & 2.607 & 0.933 & 2.299 & 0.031 & 0.003 & ---   & 1.489 & 1.800 & ---   \\
		$^{57}$Ni & 1.589 & 1.486 & 2.091 & 1.296 & 1.694 & 0.527 & 0.033 & 0.526 & 0.244 & 0.570 & 0.158 \\
		$^{61}$Zn & 1.536 & 0.795 & 1.857 & 0.605 & 1.731 & 0.446 & 0.383 & 0.367 & 2.356 & 2.888 & 3.017 \\
		$^{62}$Zn & 1.524 & 1.370 & 1.605 & 1.459 & 1.617 & 0.107 & 0.471 & 0.448 & 0.778 & 0.749 & 0.738 \\
		\hline
		\hline
	\end{tabular}
\end{table*}

\begin{table*}[]\label{T2}
	\centering
	\caption{Same as Table~1 but for C3 (upper panel) \& C4 (lower panel) cases. }
	\centering
	\begin{tabular}{c|c|cc|cc|ccc|ccc}
		\hline\hline
		&\multicolumn{5}{c}{Pairing Gaps (MeV)}	&\multicolumn{3}{c}{Total Strength (arb. units)}&\multicolumn{3}{c}{Centroids (MeV)} \\
		\hline
		Nuclei&$\bigtriangleup^{TF}_{nn=pp}$&$\bigtriangleup^{3TF}_{pp}$&$\bigtriangleup^{3TF}_{nn}$&$\bigtriangleup^{5TF}_{pp}$&$\bigtriangleup^{5TF}_{nn}$&$\sum$ GT$^{TF}$ &$\sum$ GT$^{3TF}$ &$\sum$ GT$^{5TF}$ &$\bar{E}$$^{TF}$&$\bar{E}$$^{3TF}$&$\bar{E}$$^{5TF}$\\
		\hline\hline
		$^{56}$Ni & 1.604 & 2.145 & 2.227 & 2.080 & 2.159 & 0.342 & 0.143 & 0.309 & 2.717 & 2.881 & 2.671 \\
		$^{48}$Cr & 1.732 & 2.238 & 2.228 & 2.128 & 2.135 & 0.410 & 0.432 & 0.454 & 1.135 & 1.120 & 1.151 \\
		$^{60}$Zn & 1.549 & 1.637 & 1.707 & 1.680 & 1.782 & 0.958 & 0.927 & 0.943 & 2.252 & 2.226 & 2.276 \\
		\hline
		$^{54}$Co & 1.633 & 0.859 & 0.908 & 0.967 & 1.039 & 0.740 & 0.863 & 1.168 & 5.801 & 5.632 & 6.398 \\
		$^{50}$Mn & 1.697 & 0.957 & 0.981 & 0.938 & 0.861 & 0.662 & 0.860 & 2.098 & 5.606 & 5.757 & 6.881 \\
		$^{52}$Mn & 1.664 & 0.986 & 1.165 & 1.011 & 1.160 & 0.156 & 0.082 & 0.093 & 3.797 & 3.905 & 2.773 \\
		$^{58}$Cu & 1.576 & 1.106 & 1.163 & 0.945 & 0.961 & 1.239 & 1.081 & 0.125 & 5.498 & 5.061 & 6.476 \\
		$^{64}$Co & 1.500 & 1.033 & 0.985 & 1.190 & 0.946 & 1.166 & 0.441 & 0.945 & 6.435 & 4.632 & 5.721 \\
		$^{61}$Fe & 1.536 & 1.137 & 1.423 & 1.198 & 1.417 & 0.021 & 0.028 & 0.027 & 0.004 & 0.003 & 0.003 \\
		$^{56}$Co & 1.604 & 1.212 & 1.326 & 1.349 & 1.175 & 0.101 & 0.089 & 0.040 & 3.282 & 2.531 & 3.433 \\
		$^{57}$Cr & 1.589 & 1.260 & 1.290 & 1.192 & 1.342 & 0.031 & 0.034 & 0.035 & 1.328 & 1.476 & 1.477 \\
		$^{51}$Ti & 1.680 & 1.361 & 1.503 & 1.383 & 1.224 & 0.069 & 0.050 & 0.269 & 1.127 & 1.424 & 0.032 \\
		$^{60}$Cu & 1.549 & 1.234 & 1.089 & 1.015 & 1.105 & 1.204 & 1.139 & 0.253 & 5.169 & 4.878 & 4.181 \\
		$^{62}$Cu & 1.524 & 1.217 & 1.206 & 1.067 & 1.220 & 0.005 & 0.056 & 0.058 & 3.600 & 2.931 & 3.752 \\
		$^{53}$Ti & 1.648 & 1.377 & 0.950 & 1.381 & 1.004 & 0.004 & 0.013 & 0.092 & 0.604 & 0.029 & 4.251 \\
		$^{52}$V  & 1.664 & 1.405 & 1.227 & 1.416 & 1.055 & 0.017 & 0.177 & 0.014 & 3.760 & 2.877 & 2.925 \\
		$^{55}$Cr & 1.618 & 1.402 & 1.368 & 1.311 & 1.301 & 0.196 & 0.181 & 0.175 & 0.849 & 0.936 & 0.972 \\
		$^{55}$Co & 1.618 & 1.473 & 1.170 & 1.809 & 1.248 & 0.015 & 0.009 & 0.008 & 2.720 & 3.060 & 0.405 \\
		$^{61}$Cu & 1.536 & 1.601 & 1.122 & 1.486 & 1.164 & 0.224 & ---   & ---   & 1.162 & ---   & ---   \\
		\hline\hline
	\end{tabular}
\end{table*}

\begin{table*}[]\label{T3}
	\centering
	\caption{The pn-QRPA calculated $\beta$-decay half-life values, as a function of computed pairing gap values, for  C1 (upper panel) \& C2 (lower panel) cases. The measured half-lives~\cite{Kon21} is shown in the last column.}
	
	\begin{tabular}{c|c|c|ccc|c}
		\hline\hline
		&&&\multicolumn{4}{c}{Half-lives (s)}\\
		\hline
		Nuclei&Decay Mode&Q$_\beta$&$T^{TF}_{1/2}$&$T^{3TF}_{1/2}$&$T^{5TF}_{1/2}$&$T^{Exp}_{1/2}$ \\
		\hline\hline
		$^{51}$Sc & $\beta^-$ & 6.483 & 3.88E+00 & 3.60E+00 & 7.18E+03 & 1.24E+01 \\
		$^{49}$Sc & $\beta^-$ & 2.002 & 3.43E+03 & 3.58E+03 & ---      & 3.43E+03 \\
		$^{57}$Cu & $\beta^+$ & 8.775 & 2.91E+00 & 2.05E+00 & 8.93E+00 & 1.96E-01 \\
		$^{49}$Ca & $\beta^-$ & 5.262 & 1.91E+01 & 1.90E+02 & 1.85E+02 & 5.23E+02 \\
		$^{65}$Co & $\beta^-$ & 5.941 & 1.10E+00 & 1.05E+00 & 1.43E+00 & 1.16E+00 \\
		$^{63}$Co & $\beta^-$ & 3.661 & 6.07E+00 & 2.26E+01 & 2.12E+01 & 2.69E+01 \\
		$^{62}$Fe & $\beta^-$ & 2.546 & 2.10E+02 & 8.14E+01 & 8.14E+01 & 6.80E+01 \\
		$^{58}$Cr & $\beta^-$ & 3.836 & 1.10E+01 & 1.11E+01 & 1.11E+01 & 7.00E+00 \\
		$^{59}$Cu & $\beta^+$ & 4.798 & 7.34E+01 & 7.90E+03 & 5.01E+03 & 8.15E+01 \\
		$^{59}$Mn & $\beta^-$ & 5.140 & 4.30E+00 & 4.24E+00 & 6.64E+00 & 4.59E+00 \\
		$^{57}$Mn & $\beta^-$ & 2.696 & 9.02E+01 & 8.95E+01 & 1.51E+02 & 8.54E+01 \\
		$^{50}$Sc & $\beta^-$ & 6.895 & 6.49E+01 & 4.09E+03 & 1.29E+03 & 1.03E+02 \\
		\hline
		$^{45}$Ti & $\beta^+$ & 2.062 & 3.17E+06 & 1.68E+08 & ---      & 1.11E+04 \\
		$^{57}$Ni & $\beta^+$ & 3.262 & 1.19E+02 & 2.43E+03 & 1.15E+02 & 1.28E+05 \\
		$^{61}$Zn & $\beta^+$ & 5.635 & 1.94E+01 & 2.54E+01 & 2.76E+01 & 8.91E+01 \\
		$^{62}$Zn & $\beta^+$ & 1.619 & 9.77E+04 & 2.48E+04 & 2.58E+04 & 3.31E+04 \\
		\hline\hline
	\end{tabular}
\end{table*}

\begin{table*}[]\label{T4}
	\centering
	\caption{Same as Table~3 but for C3 (upper panel) \& C4 (lower panel) cases.}
	
	\begin{tabular}{c|c|c|ccc|c}
		\hline\hline
		&&&\multicolumn{4}{c}{Half-lives (s)}\\
		\hline
		Nuclei&Decay Mode&Q$_\beta$&$T^{TF}_{1/2}$&$T^{3TF}_{1/2}$&$T^{5TF}_{1/2}$&$T^{Exp}_{1/2}$ \\
		\hline\hline
		$^{56}$Ni & $\beta^+$ & 2.133 & 2.48E+05 & 5.93E+04 & 2.22E+05 & 5.25E+05 \\
		$^{48}$Cr & $\beta^+$ & 1.657 & 1.60E+05 & 1.42E+05 & 1.46E+05 & 7.76E+04 \\
		$^{60}$Zn & $\beta^+$ & 4.171 & 1.13E+02 & 1.19E+02 & 1.21E+02 & 1.43E+02 \\
		\hline
		$^{54}$Co & $\beta^+$ & 8.245 & 4.63E+02 & 6.28E+00 & 7.62E+00 & 1.93E-01 \\
		$^{50}$Mn & $\beta^+$ & 7.634 & 6.53E+02 & 1.26E+02 & 2.76E+01 & 2.83E-01 \\
		$^{52}$Mn & $\beta^+$ & 4.708 & 1.49E+05 & 4.01E+05 & 1.23E+04 & 4.83E+05 \\
		$^{58}$Cu & $\beta^+$ & 8.561 & 5.36E+01 & 2.84E+01 & 1.28E+02 & 3.20E+00 \\
		$^{64}$Co & $\beta^-$ & 7.307 & 5.29E+00 & 1.13E+01 & 2.95E+00 & 3.00E-01 \\
		$^{61}$Fe & $\beta^-$ & 3.978 & 5.66E+01 & 4.15E+01 & 4.39E+01 & 3.59E+02 \\
		$^{56}$Co & $\beta^+$ & 4.567 & 9.46E+04 & 1.27E+04 & 3.07E+05 & 6.67E+06 \\
		$^{57}$Cr & $\beta^-$ & 4.961 & 2.51E+01 & 2.43E+01 & 2.33E+01 & 2.11E+01 \\
		$^{51}$Ti & $\beta^-$ & 2.470 & 1.90E+03 & 7.05E+03 & 4.16E+01 & 3.46E+02 \\
		$^{60}$Cu & $\beta^+$ & 6.128 & 4.44E+03 & 1.72E+03 & 6.13E+02 & 1.42E+03 \\
		$^{62}$Cu & $\beta^+$ & 3.959 & 1.89E+07 & 2.14E+05 & 2.48E+06 & 5.80E+02 \\
		$^{53}$Ti & $\beta^-$ & 4.971 & 1.48E+02 & 3.91E+01 & 4.29E+02 & 3.27E+01 \\
		$^{52}$V  & $\beta^-$ & 3.976 & 5.18E+06 & 1.56E+03 & 2.26E+04 & 2.25E+02 \\
		$^{55}$Cr & $\beta^-$ & 2.602 & 7.36E+01 & 8.24E+01 & 8.60E+01 & 2.10E+02 \\
		$^{55}$Co & $\beta^+$ & 3.451 & 3.57E+04 & 5.72E+04 & 5.82E+03 & 6.31E+04 \\
		$^{61}$Cu & $\beta^+$ & 2.238 & 7.18E+03 & ---      & ---      & 1.20E+04 \\
		\hline\hline
	\end{tabular}
\end{table*}

\begin{table*}[]\label{T5}
	\scriptsize	\caption{The state by state GT strength, branching ratios (I) and partial half-lives (in unit of $s$)  for $^{62}$Fe using the three pairing gaps (TF, 3TF and 5TF).}
	\begin{tabular}{cccc|cccc|cccc}
		\hline\hline
		\multicolumn{4}{c}{TF} & \multicolumn{4}{c}{3TF} &\multicolumn{4}{c}{5TF}\\
		\hline
		E$_x$ (MeV)& GT & I&t$^{par}_{1/2}$&E$_x$ (MeV)& GT & I&t$^{par}_{1/2}$&E$_x$ (MeV)& GT & I&t$^{par}_{1/2}$ \\
		\hline\hline
		0.143 & 0.02457 & 49.38 & 4.25E+02 & 0.081 & 0.00136 & 1.18  & 6.88E+03 & 0.081 & 0.00136 & 1.18  & 6.90E+03 \\
		0.715 & 0.00077 & 0.50  & 4.23E+04 & 0.143 & 0.04597 & 35.78 & 2.28E+02 & 0.144 & 0.04611 & 35.84 & 2.27E+02 \\
		0.794 & 0.03529 & 18.97 & 1.11E+03 & 0.184 & 0.00001 & 0.01  & 1.59E+06 & 0.184 & 0.00001 & 0.00  & 1.86E+06 \\
		0.905 & 0.03978 & 16.36 & 1.28E+03 & 0.221 & 0.01682 & 11.38 & 7.16E+02 & 0.221 & 0.01685 & 11.39 & 7.14E+02 \\
		0.988 & 0.00423 & 1.41  & 1.49E+04 & 0.362 & 0.06835 & 35.54 & 2.29E+02 & 0.363 & 0.06842 & 35.52 & 2.29E+02 \\
		1.141 & 0.00545 & 1.20  & 1.75E+04 & 0.465 & 0.00494 & 2.10  & 3.87E+03 & 0.465 & 0.00494 & 2.10  & 3.88E+03 \\
		1.169 & 0.01008 & 2.05  & 1.03E+04 & 0.650 & 0.00092 & 0.27  & 3.05E+04 & 0.651 & 0.00094 & 0.27  & 3.00E+04 \\
		1.243 & 0.01578 & 2.58  & 8.15E+03 & 0.701 & 0.00091 & 0.24  & 3.47E+04 & 0.702 & 0.00089 & 0.23  & 3.56E+04 \\
		1.312 & 0.03945 & 5.20  & 4.04E+03 & 0.812 & 0.00021 & 0.04  & 1.90E+05 & 0.813 & 0.00021 & 0.04  & 1.92E+05 \\
		1.425 & 0.00390 & 0.35  & 5.94E+04 & 0.869 & 0.00004 & 0.01  & 1.26E+06 & 0.870 & 0.00003 & 0.01  & 1.34E+06 \\
		1.539 & 0.00283 & 0.17  & 1.24E+05 & 0.948 & 0.01186 & 1.70  & 4.79E+03 & 0.949 & 0.01202 & 1.72  & 4.74E+03 \\
		1.621 & 0.00394 & 0.17  & 1.23E+05 & 1.065 & 0.00018 & 0.02  & 4.33E+05 & 1.066 & 0.00019 & 0.02  & 4.03E+05 \\
		1.637 & 0.00078 & 0.03  & 6.61E+05 & 1.110 & 0.00174 & 0.16  & 5.00E+04 & 1.111 & 0.00175 & 0.16  & 5.01E+04 \\
		1.675 & 0.01025 & 0.36  & 5.92E+04 & 1.121 & 0.01264 & 1.14  & 7.13E+03 & 1.122 & 0.01285 & 1.16  & 7.03E+03 \\
		1.701 & 0.00207 & 0.06  & 3.29E+05 & 1.154 & 0.00015 & 0.01  & 6.67E+05 & 1.156 & 0.00014 & 0.01  & 6.95E+05 \\
		1.750 & 0.01746 & 0.43  & 4.86E+04 & 1.155 & 0.00207 & 0.17  & 4.80E+04 & 1.157 & 0.00197 & 0.16  & 5.06E+04 \\
		1.774 & 0.00098 & 0.02  & 9.75E+05 & 1.276 & 0.00357 & 0.20  & 3.98E+04 & 1.277 & 0.00361 & 0.21  & 3.95E+04 \\
		1.777 & 0.00062 & 0.01  & 1.57E+06 & 1.289 & 0.00452 & 0.25  & 3.27E+04 & 1.290 & 0.00442 & 0.24  & 3.36E+04 \\
		1.815 & 0.03983 & 0.72  & 2.92E+04 & 1.292 & 0.00571 & 0.31  & 2.62E+04 & 1.293 & 0.00567 & 0.31  & 2.65E+04 \\
		2.035 & 0.00050 & 0.00  & 8.40E+06 & 1.329 & 0.04499 & 2.18  & 3.73E+03 & 1.330 & 0.04506 & 2.17  & 3.74E+03 \\
		2.162 & 0.01347 & 0.03  & 8.43E+05 & 1.365 & 0.09366 & 4.04  & 2.02E+03 & 1.366 & 0.09385 & 4.02  & 2.02E+03 \\
		---   & ---     & ---   & ---      & 1.408 & 0.01079 & 0.40  & 2.02E+04 & 1.409 & 0.01090 & 0.40  & 2.01E+04 \\
		---   & ---     & ---   & ---      & 1.459 & 0.05561 & 1.73  & 4.70E+03 & 1.461 & 0.05558 & 1.72  & 4.72E+03 \\
		---   & ---     & ---   & ---      & 1.489 & 0.00079 & 0.02  & 3.70E+05 & 1.490 & 0.00078 & 0.02  & 3.72E+05 \\
		---   & ---     & ---   & ---      & 1.492 & 0.00555 & 0.15  & 5.30E+04 & 1.494 & 0.00556 & 0.15  & 5.32E+04 \\
		---   & ---     & ---   & ---      & 1.551 & 0.00008 & 0.00  & 4.45E+06 & 1.552 & 0.00008 & 0.00  & 4.89E+06 \\
		---   & ---     & ---   & ---      & 1.632 & 0.00301 & 0.05  & 1.68E+05 & 1.633 & 0.00300 & 0.05  & 1.69E+05 \\
		---   & ---     & ---   & ---      & 1.688 & 0.00019 & 0.00  & 3.39E+06 & 1.689 & 0.00019 & 0.00  & 3.46E+06 \\
		---   & ---     & ---   & ---      & 1.711 & 0.00022 & 0.00  & 3.25E+06 & 1.713 & 0.00022 & 0.00  & 3.28E+06 \\
		---   & ---     & ---   & ---      & 1.713 & 0.04048 & 0.46  & 1.77E+04 & 1.714 & 0.04057 & 0.46  & 1.78E+04 \\
		---   & ---     & ---   & ---      & 1.726 & 0.00246 & 0.03  & 3.09E+05 & 1.728 & 0.00243 & 0.03  & 3.15E+05 \\
		---   & ---     & ---   & ---      & 1.840 & 0.00176 & 0.01  & 7.49E+05 & 1.841 & 0.00177 & 0.01  & 7.50E+05 \\
		---   & ---     & ---   & ---      & 1.843 & 0.00505 & 0.03  & 2.65E+05 & 1.844 & 0.00505 & 0.03  & 2.67E+05 \\
		---   & ---     & ---   & ---      & 1.874 & 0.02314 & 0.12  & 6.83E+04 & 1.875 & 0.02327 & 0.12  & 6.85E+04 \\
		---   & ---     & ---   & ---      & 1.938 & 0.03372 & 0.12  & 6.72E+04 & 1.939 & 0.03374 & 0.12  & 6.77E+04 \\
		---   & ---     & ---   & ---      & 2.000 & 0.03030 & 0.07  & 1.10E+05 & 2.002 & 0.03031 & 0.07  & 1.11E+05 \\
		---   & ---     & ---   & ---      & 2.121 & 0.00409 & 0.00  & 1.96E+06 & 2.122 & 0.00405 & 0.00  & 2.01E+06 \\
		---   & ---     & ---   & ---      & 2.123 & 0.00799 & 0.01  & 1.02E+06 & 2.124 & 0.00805 & 0.01  & 1.02E+06 \\
		---   & ---     & ---   & ---      & 2.222 & 0.09247 & 0.04  & 2.20E+05 & 2.224 & 0.09249 & 0.04  & 2.23E+05 \\
		---   & ---     & ---   & ---      & 2.289 & 0.05131 & 0.01  & 8.56E+05 & 2.291 & 0.05134 & 0.01  & 8.72E+05 \\
		\hline\hline
	\end{tabular}
\end{table*}

\begin{table*}[]\label{T6}
	\scriptsize	\caption{Same as Table~5, but for $^{62}$Zn.}
	\begin{tabular}{cccc|cccc|cccc}
		\hline\hline
		\multicolumn{4}{c}{TF} & \multicolumn{4}{c}{3TF} &\multicolumn{4}{c}{5TF}\\
		\hline
		E$_x$ (MeV)& GT & I&t$^{par}_{1/2}$&E$_x$ (MeV)& GT & I&t$^{par}_{1/2}$&E$_x$ (MeV)& GT & I&t$^{par}_{1/2}$ \\
		\hline\hline
		0.001 & 0.00101 & 2.87  & 3.40E+06 & 0.103 & 0.00110 & 0.63  & 3.94E+06 & 0.103 & 0.00097 & 0.58  & 4.47E+06 \\
		0.020 & 0.00265 & 7.22  & 1.35E+06 & 0.155 & 0.00794 & 4.08  & 6.07E+05 & 0.154 & 0.00756 & 4.05  & 6.36E+05 \\
		0.045 & 0.02383 & 61.31 & 1.59E+05 & 0.207 & 0.04305 & 19.98 & 1.24E+05 & 0.205 & 0.04335 & 20.94 & 1.23E+05 \\
		0.469 & 0.00041 & 0.47  & 2.07E+07 & 0.279 & 0.11067 & 44.86 & 5.53E+04 & 0.277 & 0.10029 & 42.38 & 6.08E+04 \\
		0.482 & 0.00058 & 0.65  & 1.50E+07 & 0.280 & 0.02153 & 8.72  & 2.84E+05 & 0.279 & 0.01902 & 8.02  & 3.21E+05 \\
		0.511 & 0.01289 & 13.71 & 7.13E+05 & 0.565 & 0.00002 & 0.01  & 4.41E+08 & 0.574 & 0.00001 & 0.00  & 9.98E+08 \\
		0.785 & 0.00213 & 1.28  & 7.64E+06 & 0.593 & 0.00009 & 0.02  & 1.13E+08 & 0.602 & 0.00003 & 0.01  & 3.86E+08 \\
		0.810 & 0.00645 & 3.63  & 2.69E+06 & 0.607 & 0.00345 & 0.78  & 3.20E+06 & 0.615 & 0.00239 & 0.55  & 4.70E+06 \\
		1.106 & 0.00361 & 0.81  & 1.21E+07 & 0.669 & 0.00085 & 0.17  & 1.48E+07 & 0.669 & 0.00044 & 0.09  & 2.82E+07 \\
		1.137 & 0.00553 & 1.09  & 8.93E+06 & 0.694 & 0.00084 & 0.16  & 1.58E+07 & 0.698 & 0.00075 & 0.14  & 1.79E+07 \\
		1.204 & 0.04768 & 6.96  & 1.40E+06 & 0.699 & 0.00051 & 0.09  & 2.64E+07 & 0.703 & 0.00736 & 1.41  & 1.83E+06 \\
		1.603 & 0.00040 & 0.00  & 3.48E+11 & 0.700 & 0.00161 & 0.30  & 8.33E+06 & 0.716 & 0.02746 & 5.09  & 5.06E+05 \\
		---  & ---     & ---   & ---      & 0.747 & 0.03219 & 5.36  & 4.63E+05 & 0.762 & 0.00247 & 0.41  & 6.26E+06 \\
		---  & ---     & ---   & ---      & 0.758 & 0.00087 & 0.14  & 1.75E+07 & 0.881 & 0.00793 & 0.98  & 2.63E+06 \\
		---  & ---     & ---   & ---      & 0.878 & 0.00896 & 1.07  & 2.31E+06 & 0.898 & 0.00004 & 0.01  & 5.29E+08 \\
		---  & ---     & ---   & ---      & 0.895 & 0.00004 & 0.01  & 4.90E+08 & 0.924 & 0.08483 & 9.26  & 2.78E+05 \\
		---  & ---     & ---   & ---      & 0.945 & 0.06869 & 6.80  & 3.65E+05 & 1.018 & 0.00006 & 0.01  & 5.24E+08 \\
		---  & ---     & ---   & ---      & 1.014 & 0.00005 & 0.00  & 5.88E+08 & 1.102 & 0.07279 & 4.37  & 5.90E+05 \\
		---  & ---     & ---   & ---      & 1.108 & 0.07320 & 4.12  & 6.01E+05 & 1.135 & 0.00037 & 0.02  & 1.32E+08 \\
		---  & ---     & ---   & ---      & 1.125 & 0.00037 & 0.02  & 1.26E+08 & 1.165 & 0.02261 & 1.04  & 2.47E+06 \\
		---  & ---     & ---   & ---      & 1.162 & 0.04401 & 1.98  & 1.25E+06 & 1.211 & 0.01113 & 0.41  & 6.24E+06 \\
		---  & ---     & ---   & ---      & 1.204 & 0.01179 & 0.44  & 5.68E+06 & 1.398 & 0.01931 & 0.20  & 1.27E+07 \\
		---  & ---     & ---   & ---      & 1.399 & 0.02065 & 0.21  & 1.19E+07 & 1.466 & 0.00453 & 0.02  & 1.16E+08 \\
		---  & ---     & ---   & ---      & 1.463 & 0.00521 & 0.03  & 9.76E+07 & 1.513 & 0.01245 & 0.03  & 9.33E+07 \\
		---  & ---     & ---   & ---      & 1.511 & 0.01321 & 0.03  & 8.43E+07 & ---   & ---     & ---   & ---      \\
		\hline\hline
	\end{tabular}
\end{table*}

\begin{table*}[]\label{T7}
	\scriptsize	\caption{Same as Table~5, but for $^{48}$Cr.}
	\begin{tabular}{cccc|cccc|cccc}
		\hline\hline
		\multicolumn{4}{c}{TF} & \multicolumn{4}{c}{3TF} &\multicolumn{4}{c}{5TF}\\
		\hline
		E$_x$ (MeV)& GT & I&t$^{par}_{1/2}$&E$_x$ (MeV)& GT & I&t$^{par}_{1/2}$&E$_x$ (MeV)& GT & I&t$^{par}_{1/2}$ \\
		\hline\hline
		0.430 & 0.01244 & 12.39 & 1.29E+06 & 0.383 & 0.00694 & 6.72  & 2.11E+06 & 0.395 & 0.00802 & 7.81  & 1.87E+06 \\
		0.583 & 0.00081 & 0.61  & 2.61E+07 & 0.468 & 0.00061 & 0.50  & 2.82E+07 & 0.494 & 0.00062 & 0.50  & 2.91E+07 \\
		0.607 & 0.00237 & 1.71  & 9.34E+06 & 0.560 & 0.00283 & 1.98  & 7.16E+06 & 0.582 & 0.00725 & 4.99  & 2.92E+06 \\
		0.676 & 0.00064 & 0.40  & 3.96E+07 & 0.584 & 0.01844 & 12.30 & 1.15E+06 & 0.594 & 0.00708 & 4.76  & 3.06E+06 \\
		0.762 & 0.01107 & 5.78  & 2.76E+06 & 0.702 & 0.09195 & 48.62 & 2.91E+05 & 0.720 & 0.10358 & 54.08 & 2.69E+05 \\
		0.798 & 0.13156 & 63.26 & 2.52E+05 & 0.711 & 0.00655 & 3.39  & 4.18E+06 & 0.724 & 0.00742 & 3.85  & 3.78E+06 \\
		0.987 & 0.00005 & 0.01  & 1.16E+09 & 0.873 & 0.00026 & 0.09  & 1.55E+08 & 0.898 & 0.00020 & 0.07  & 2.14E+08 \\
		1.105 & 0.04450 & 8.79  & 1.82E+06 & 0.996 & 0.04170 & 10.50 & 1.35E+06 & 1.021 & 0.04356 & 10.43 & 1.40E+06 \\
		1.260 & 0.00023 & 0.02  & 6.82E+08 & 1.135 & 0.01982 & 3.10  & 4.57E+06 & 1.164 & 0.01410 & 2.02  & 7.19E+06 \\
		1.302 & 0.00893 & 0.72  & 2.22E+07 & 1.181 & 0.05302 & 6.89  & 2.06E+06 & 1.212 & 0.04554 & 5.32  & 2.74E+06 \\
		1.409 & 0.08462 & 3.29  & 4.85E+06 & 1.317 & 0.02635 & 1.74  & 8.16E+06 & 1.334 & 0.03692 & 2.25  & 6.47E+06 \\
		1.450 & 0.10984 & 2.96  & 5.39E+06 & 1.369 & 0.06005 & 2.81  & 5.03E+06 & 1.385 & 0.07115 & 3.06  & 4.75E+06 \\
		1.497 & 0.00322 & 0.05  & 3.11E+08 & 1.380 & 0.00069 & 0.03  & 4.76E+08 & 1.406 & 0.00110 & 0.04  & 3.64E+08 \\
		---   &  ---    & ---   &   ---    & 1.503 & 0.10312 & 1.33  & 1.06E+07 & 1.539 & 0.10725 & 0.82  & 1.77E+07 \\
		\hline\hline
	\end{tabular}
\end{table*}

\begin{table*}[]\label{T8}
	\scriptsize	\caption{Same as Table~5, but for $^{58}$Cu.}
	\begin{tabular}{cccc|cccc|cccc}
		\hline\hline
		\multicolumn{4}{c}{TF} & \multicolumn{4}{c}{3TF} &\multicolumn{4}{c}{5TF}\\
		\hline
		E$_x$ (MeV)& GT & I&t$^{par}_{1/2}$&E$_x$ (MeV)& GT & I&t$^{par}_{1/2}$&E$_x$ (MeV)& GT & I&t$^{par}_{1/2}$ \\
		\hline\hline
		4.353 & 0.01931 & 9.17  & 5.84E+02 & 3.743 & 0.01123 & 6.21  & 4.57E+02 & 2.908 & 0.00098 & 6.10  & 2.10E+03 \\
		4.353 & 0.01931 & 9.17  & 5.84E+02 & 3.743 & 0.01123 & 6.21  & 4.57E+02 & 2.908 & 0.00098 & 6.10  & 2.10E+03 \\
		5.302 & 0.22210 & 22.34 & 2.40E+02 & 4.872 & 0.00142 & 0.16  & 1.75E+04 & 3.568 & 0.01328 & 40.73 & 3.15E+02 \\
		5.302 & 0.22210 & 22.34 & 2.40E+02 & 4.872 & 0.00142 & 0.16  & 1.75E+04 & 3.568 & 0.01328 & 40.73 & 3.15E+02 \\
		5.396 & 0.00283 & 0.24  & 2.26E+04 & 4.874 & 0.21000 & 23.95 & 1.18E+02 & 3.656 & 0.00003 & 0.09  & 1.43E+05 \\
		5.396 & 0.00283 & 0.24  & 2.26E+04 & 4.874 & 0.21000 & 23.95 & 1.18E+02 & 3.656 & 0.00003 & 0.09  & 1.43E+05 \\
		5.629 & 0.28747 & 14.81 & 3.62E+02 & 5.198 & 0.04107 & 2.66  & 1.07E+03 & 4.725 & 0.00081 & 0.53  & 2.41E+04 \\
		5.629 & 0.28747 & 14.81 & 3.62E+02 & 5.198 & 0.04107 & 2.66  & 1.07E+03 & 4.725 & 0.00081 & 0.53  & 2.41E+04 \\
		5.707 & 0.07852 & 3.40  & 1.57E+03 & 5.219 & 0.27226 & 16.98 & 1.67E+02 & 4.754 & 0.00274 & 1.71  & 7.49E+03 \\
		5.707 & 0.07852 & 3.40  & 1.57E+03 & 5.219 & 0.27226 & 16.98 & 1.67E+02 & 4.754 & 0.00274 & 1.71  & 7.49E+03 \\
		6.532 & 0.00754 & 0.04  & 1.50E+05 & 6.054 & 0.00363 & 0.04  & 7.92E+04 & 5.056 & 0.00017 & 0.06  & 2.00E+05 \\
		6.532 & 0.00754 & 0.04  & 1.50E+05 & 6.054 & 0.00363 & 0.04  & 7.92E+04 & 5.056 & 0.00017 & 0.06  & 2.00E+05 \\
		7.896 & 0.00182 & 0.00  & 1.60E+07 & 7.449 & 0.00092 & 0.00  & 1.13E+07 & 5.104 & 0.00202 & 0.70  & 1.82E+04 \\
		7.896 & 0.00182 & 0.00  & 1.60E+07 & 7.449 & 0.00092 & 0.00  & 1.13E+07 & 5.104 & 0.00202 & 0.70  & 1.82E+04 \\
		---   & ---     & ---   & ---      &  ---  & ---     & ---   &  ---     & 5.921 & 0.00036 & 0.02  & 5.65E+05 \\
		---   & ---     & ---   & ---      &  ---  & ---     & ---   &  ---     & 5.921 & 0.00036 & 0.02  & 5.65E+05 \\
		---   & ---     & ---   & ---      &  ---  & ---     & ---   &  ---     & 7.094 & 0.01177 & 0.03  & 4.55E+05 \\
		---   & ---     & ---   & ---      &  ---  & ---     & ---   &  ---     & 7.173 & 0.01149 & 0.02  & 5.47E+05 \\
		---   & ---     & ---   & ---      &  ---  & ---     & ---   &  ---     & 7.173 & 0.01149 & 0.02  & 5.47E+05 \\
		---   & ---     & ---   & ---      &  ---  & ---     & ---   &  ---     & 7.324 & 0.00088 & 0.00  & 9.41E+06 \\
		---   & ---     & ---   & ---      &  ---  & ---     & ---   &  ---     & 7.324 & 0.00088 & 0.00  & 9.41E+06 \\
		---   & ---     & ---   & ---      &  ---  & ---     & ---   &  ---     & 7.603 & 0.00314 & 0.00  & 4.44E+06 \\
		---   & ---     & ---   & ---      &  ---  & ---     & ---   &  ---     & 7.856 & 0.00000 & 0.00  & 1.30E+10 \\
		---   & ---     & ---   & ---      &  ---  & ---     & ---   &  ---     & 8.062 & 0.00756 & 0.00  & 6.88E+06 \\
		---   & ---     & ---   & ---      &  ---  & ---     & ---   &  ---     & 8.169 & 0.03616 & 0.01  & 2.35E+06 \\
		---   & ---     & ---   & ---      &  ---  & ---     & ---   &  ---     & 8.221 & 0.00002 & 0.00  & 6.60E+09 \\
		---   & ---     & ---   & ---      &  ---  & ---     & ---   &  ---     & 8.264 & 0.00017 & 0.00  & 8.56E+08 \\
		---   & ---     & ---   & ---      &  ---  & ---     & ---   &  ---     & 8.412 & 0.00025 & 0.00  & 2.53E+09 \\
		---   & ---     & ---   & ---      &  ---  & ---     & ---   &  ---     & 8.420 & 0.00038 & 0.00  & 1.87E+09 \\
		\hline
		\hline
	\end{tabular}
\end{table*}


\end{document}